\documentclass[12pt]{iopart}
\usepackage{graphicx}
\usepackage{multicol}
\begin{document}

\title{Optimal generation of Fock states in a weakly nonlinear oscillator}

\author{B. Khani, J.M. Gambetta, F. Motzoi, and F.K. Wilhelm}
\address{Institute for Quantum Computing and Department of Physics and
Astronomy, University of Waterloo, 200 University Avenue West,
Waterloo, ON, Canada, N2L 3G1}
\ead{fwilhelm@iqc.ca}
\begin{abstract}
We apply optimal control theory to determine the shortest time in which an energy eigenstate of a weakly anharmonic oscillator can be created under the practical constraint of linear driving. We show that the optimal pulses are beatings of mostly the transition frequencies for the transitions up to the desired state and the next leakage level. The time of a shortest possible pulse for a given nonlinearity scale with the nonlinearity parameter $\delta$ as a power law $\delta^{-\alpha}$ with $\alpha\simeq 0.73\pm0.029$.  This is a qualitative improvement relative to the value $\alpha=1$ suggested by a simple Landau-Zener argument.
\end{abstract}
\maketitle

\section{Introduction}

Physics is fundamentally quantum. On the other hand, in practice, many objects and processes appear to follow the laws of classical physics. This includes objects coupled to large baths with manydegrees of freedom, 
hence suffering from decoherence. Furthermore, in objects of large mass, quantum phenomena occur at an unobservably short de Broglie wavelength \cite{Leggett02}. Harmonic oscillators, on the other hand, even if they are only weakly damped and have low mass and thus must be in the quantum regime, may appear close to classical: These linear systems are semiclassical, meaning that in their ground state and thermal equilibrium state, as well as in any states created from these by an arbitrary linear drive, correspond to the classical limit small quantum fluctuations, essentially following Ehrenfest's theorem. The smallness  of these fluctuations and their competition with thermal fluctuations makes it hard to verify the quantum nature of these systems. A way out is to add an extra nonlinear ingredient to create a nonclassical state, such as a Fock state (an energy eigenstate of the oscillator) or a squeezed state (a state with quantum fluctuations in one variable below the standard Heisenberg limit). Examples include coupling the oscillator to a nonlinear object such as a qubit \cite{Houck07,Hofheinz08,Bertet02,Law96,Blencowe04,Naik06}. In this work, we focus on a different approach, which is exploiting a small nonlinearity of the oscillator
potential. These nonlinearities occur, e.g., in nanomechanical systems \cite{Cleland03,Kozinsky07}, Josephson circuits \cite{Yurke87,Siddiqi04,Lupascu07,Ioana07c,Dykman09}, optics \cite{Walls94,Collett84,Collett85}, ion traps \cite{Brown86,Alheit97}, and molecules \cite{Tesch02}. These systems that are rather large in size on an atomic scale are in particular important in this context, 
because this would enable to show the quantum character of center-of-mass variables of mesos- and macroscopic objects. Also, Fock states are useful for the demonstration
of short-range quantum communication on a chip.

The plan of the paper is as follows: we are first going to define our nonlinear oscillator Hamiltonian and outline the difficulty of creating Fock states in this system, as well as the GRAPE method used to find optimized ways to prepare these Fock states. We will then present optimized pulses, identifying a beating structure that can easily be connected to the energy spectrum. We are finally discussing the intrinsic speed limits for Fock state preparation.

\section{Setting and approach}

\subsection{Hamiltonian \label{ch:Hamiltonian}}

\begin{figure}
\includegraphics[width=\columnwidth]{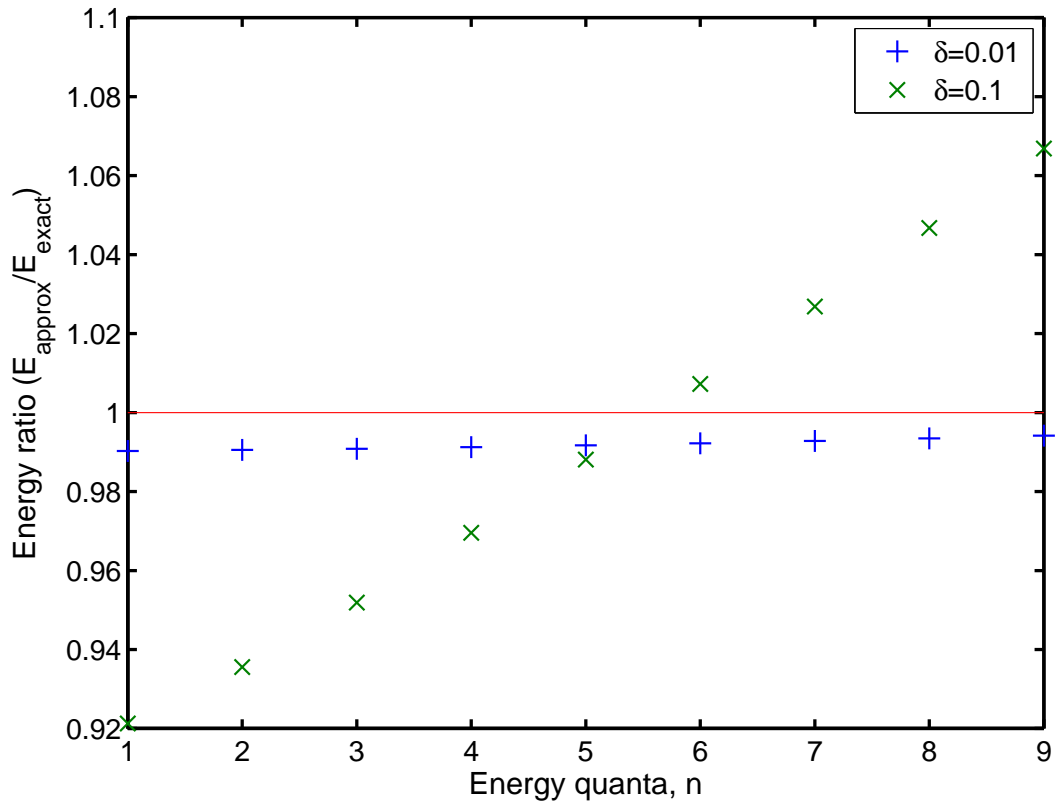}
\caption{\label{fig:spectrum} Ratio of the eigenergies of the RWA Hamiltonian, eq. \ref{eq:RWAhamiltonian} and  exact eigenergies of the anharmonic oscillator for the first nine energy levels, assuming nonlinearities $\delta=0.01$ and $\delta=0.1$.}
\end{figure}

We study a weakly nonlinear oscillator, the Duffing oscillator, with Hamiltonian
\begin{equation}
H=\frac{p^2}{2m}+\frac{m\omega_0^2}{2}x^2+\delta \frac{m^2\omega_0^2}{3\hbar}x^4\quad \delta>0
\end{equation}
where $x$ and $p$ are coordinate and momentum, respectively, where $\omega_0$ is the resonance frequency in the limit $\delta\rightarrow0$.
Even though we focus on the hard case, $\delta>0$, we expect similar conclusions to hold for the soft case. 
This Hamiltonian can be made dimensionless with raising and lowering operators analogous to the harmonic oscillator \cite{Cohen92}. It then reads
\begin{equation}
H=\hbar\omega_0\left(a^\dagger a+\frac{1}{2}\right)+\frac{\hbar\delta}{12}(a+a^\dagger)^4. 
\end{equation}
By expanding the nonlinearity and ordering its terms, we can split the Hamiltonian 
as $H=H_0(\hat{n})+\hat{V}$ into a terms $H_0$ that contains only the number operator $\hat{n}=a^\dagger a$ that will dominate perturbation theory for small $\delta/\hbar\omega\ll 1$ and corresponds to the rotating wave approximation, and a term $\hat{V}$ that goes beyond that approximation. We find 
\begin{eqnarray}
H_0(\hat{n})&=\hbar\omega_0\left(\hat{n}+\frac{1}{2}\right)+\frac{\hbar\delta}{4}(2\hat{n}^2+2\hat{n}+1).& \\
V&=\frac{\delta}{12}(a^{\dagger 4}+6a^{\dagger 2}+4a^\dagger(a^2 + a^{\dagger 2})a + 6a^2 + a^4).&
\label{eq:V}
\end{eqnarray}
$H_0$ can be re-parameterized to a simpler form using $\omega=\omega_0+\delta$,  while dropping a constant energy shift
\begin{eqnarray}
H_0&=\hbar\left[\omega \hat{n}+\frac{1}{2}\delta{\hat{n}}(\hat{n}-1)\right]. &\label{eq:RWAhamiltonian}
\end{eqnarray}
where $\omega$ is the frequency of the $|0\rangle\leftrightarrow|1\rangle$ transition.
$H$ can be diagonalized straightforwardly and its eigenstates $|n\rangle$, $n=0,1,2, \dots$ are close to the harmonic oscillator eigenstates, see fig. \ref{fig:spectrum}, at small $\delta$ and the eigenenergies are given by the polynomial deriving from eq. (\ref{eq:RWAhamiltonian}), $E_n\simeq H_0(n)$. We can note that for small $\delta$, the energy splittings between adjacent levels are only slightly different from each other. Denoting 
the transition frequencies as $\omega_{ij}=(E_i-E_j)/\hbar$, this means that 
$\omega_{n+1,n}=\omega+\delta n$ only has a weak $n$ dependence by $\delta\ll\omega$. 

In the most common physical situation, external driving is given by
\begin{equation}
H_c(t)=F(t)x=\hbar f(t)(a+a^\dagger)
\label{eq:controlham}
\end{equation}
corresponding to current drive in the Josephson case, a laser field in the ion trap case and an external force in the nanomechanical case. Here, $f(t)$ is proportional to the force $F$ \cite{Cohen92}. Given that the eigenstates are almost harmonic oscillator eigenstates, this drive predominantly, in the sense of an approximate selection rule, couples adjacent levels, $n$ and $n+1$. 

This setting outlines two difficulties in the task of preparing Fock states starting from the ground state using a straightforward application of $H_c$. Firstly, there is no direct transition from the ground state to any state higher than $|1\rangle$ unless $\delta$ is large enough to make the eigenstate corrections due to $\hat{V}$, eq. (\ref{eq:V}), significant. Secondly, even for preparing the first excited state, the approximate Liouvillian degeneracy (i.e., the weak $n$-dependence of $\omega_{n+1,n}$),  prohibits short pulses: A short resonant $\pi$-pulse of length $T_r$
has, by the energy-time uncertainty principle of Fourier transform, a bandwidth of $\omega_B\simeq\pi/T_r$. If $\omega_B$ becomes comparable to the small
parameter $|\omega_{n+1,n}-\omega_{n,n-1}|=\delta$,  a pulse applied to $\omega_{n,n-1}$ will have a significant Fourier component at $\omega_{n+1,n}$.  Hence, driving the transition from $n-1$ to $n$ will inevitably also drive $n$ to $n+1$ inducing leakage to the next higher state $n+1$. This could be overcome by a very long $\pi$-pulse that keeps $\omega_B$ small, however, in a realistic setting, this will compete with energy relaxation back to the ground state at a rate $T_1$, effectively limiting $\omega_B\gg 1/T_1$.

\subsection{GRAPE}

Given this conundrum, we resort to optimal control to find a way to prepare higher Fock states. We use the GRAPE algorithm \cite{Khaneja05}. This
is an optimal
control algorithm that uses open-loop control for optimizing control fields. The
algorithm is: given an initial state $|\psi_i\rangle$, a target state $|\psi_f\rangle$ and a
time period for the state transfer $T$, such that $|\psi(t_0)\rangle=|\psi_i\rangle$ and 
$|\psi(t_0+T)\rangle=|\psi_f\rangle$.  The search starts from an initial guess for the controls, which for us is always taken to be a constant-envelope resonant $\pi$-pulse 
with frequency $\omega$, i.e., $f(t)=\Omega_R\cos\omega t$. $\Omega_R$ turns out to be the Rabi frequency of the $0\leftrightarrow 1$ transition. The time $T$ is sliced into $N$ intervals of length
$\delta t=T/N$ with $N$ between 101 and 401 and pulse amplitudes are assumed to 
be constant across the intervals. 
As a performance index, we use 
\begin{equation}
\Phi=|\left\langle \psi_i | \psi_f\right\rangle|^2.
\end{equation}
If $f_i$ designates the value of the control field $f$ in eq. (\ref{eq:controlham}) in time step $i$, it can be shown that
\begin{equation}
\frac{\partial\Phi}{\partial f_j}=-i\delta t\left\langle \lambda_j\left[a+a^\dagger,\rho_j\right]\right\rangle.
\label{eq:gradient}
\end{equation}
Here, $\rho_j$ is the projector on the state in time step $j$ and we see 
the structure of eq. (\ref{eq:controlham}). Here $\lambda_j=|\psi_j\rangle\langle \psi_j|$ is the projector on the back-propagated target state
\begin{equation}
|\psi_j\rangle=U_{j+1}^\dagger\cdots U_N^\dagger |\psi_f\rangle.
\end{equation}
Here, $U_i$ is the propagator for the Schr\"odinger equation across time step $j$.
The gradient eq. (\ref{eq:gradient}) can be very efficiently calculated this way and 
used for a gradient search for a locally optimal set of values $f_i$ describing the desired pulse shape.  We temporarily increased the gradient by a  factor of two for each time the pulse was found to be in an undesired local optimum.

\section{Numerical results}

\begin{figure}
\includegraphics[width=15cm]{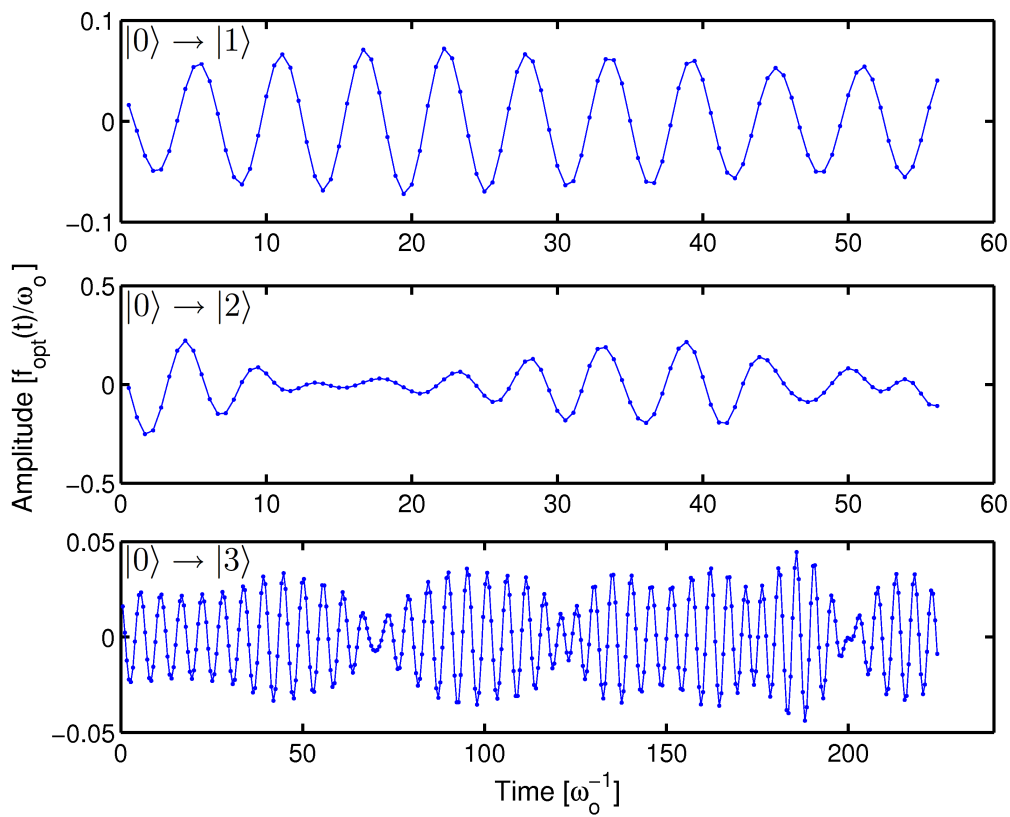}
\caption{\label{fig:pulses} The amplitudes of three pulse sequences optimized by GRAPE for preparation of $|1\rangle$ (top), $|2\rangle$ (middle), and $|3\rangle$ (bottom) are shown. 
Each system had a Hilbert space consisting of 10 eigenstates, with $\delta=0.12$.  Each optimized pulse leads to a fidelity of 0.9999.}
\end{figure}

\begin{figure}
\includegraphics[width=15cm]{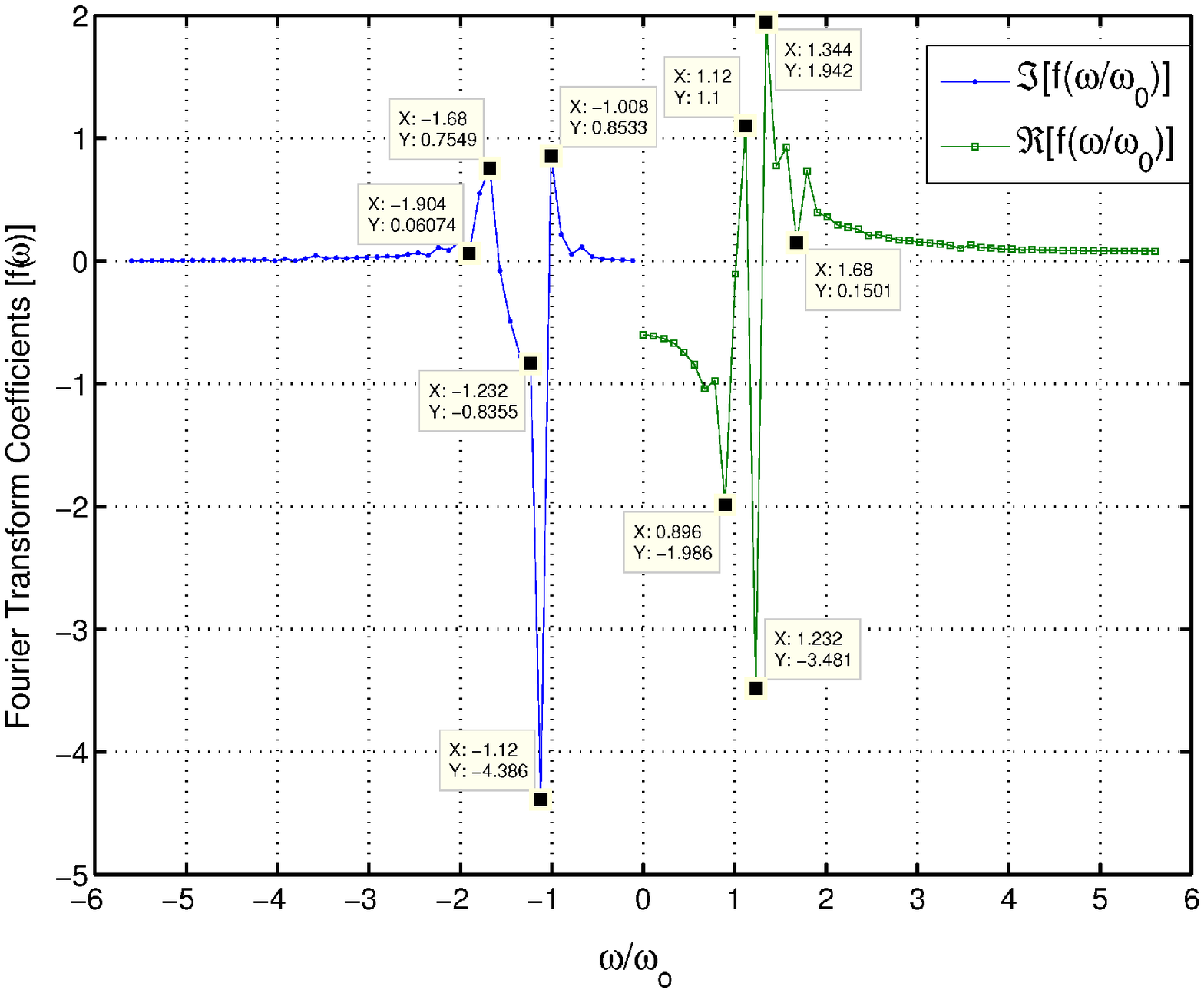}
\caption{\label{fig:spectra} Fourier specrum for preparing $|2\rangle$.  To the left and right of the zero frequency is the imaginary and real coefficients, respectively, of the decomposition.  The main peaks in each pulse spectrum is labeled.  The transition frequencies between $|0\rangle$ and $|1\rangle$ ($\omega=1.12$) are apparent in each pulse.  The transition frequency $\omega_{21}=1.24$ and $\omega_{32}=1.36$ is apparent in the pulse preparing the $|2\rangle$ and $|3\rangle$ states.}  
\end{figure}

We have analyzed this setting in a number of cases. In all cases, the initial state was chosen to be the ground state, $|\psi_i\rangle=|0\rangle$ and the first, second, and third excited state where chosen as final states, $|\psi_f\rangle=|n=1,2,3\rangle$. We have truncated the energy spectrum at $n=10$ levels and verified, that changing the number of levels did not lead to a discernible change in the pulse. In fig. \ref{fig:pulses} we show optimized pulses that display clear beating behavior. 
This beating becomes more complex as we go to higher states. More clearly, this structure can be analyzed by its Fourier transform, as shown in fig. \ref{fig:spectra}.  
We observe a cluster of close discrete lines, confirming the observation that the pulses are beatings. The predominant frequencies in the $|0\rangle\rightarrow |2\rangle$ transition are $\omega_{10}=1.12$, the slightly higher $\omega_{21}=1.24$ and $\omega_{32}=1.36$, with additional sidebands. Analyzing more transitions, we see the same picture, where a transition from $|0\rangle$ to $|n\rangle$ contains frequencies $\omega_{m+1,m}$ for 
$0\le m \le n$. 

This structure can be understood as follows: Due to the oscillator selection rule, the ladder of states needs to be climbed sequentially, it is not possible to go from $|n\rangle$ to $|n+k\rangle$ for any $k>1$, explaining the frequencies $\omega_{10}$ through 
$\omega_{n,n-1}$ for a transition $0\rightarrow n$ that correspond to driving $\pi$-pulses on those respective transitions.  Due to the bandwith issue discussed in section \ref{ch:Hamiltonian}, this also drives the transition to level $n+1$, it is also necessary to drive the leakage transition leading out of $|n\rangle$, i.e., at $\omega_{n,n+1}$. A similar effect is seen in quantum gates for three-state systems \cite{Rebentrost09,Motzoi09,Safaei09}

\begin{figure}
\includegraphics[width=15cm]{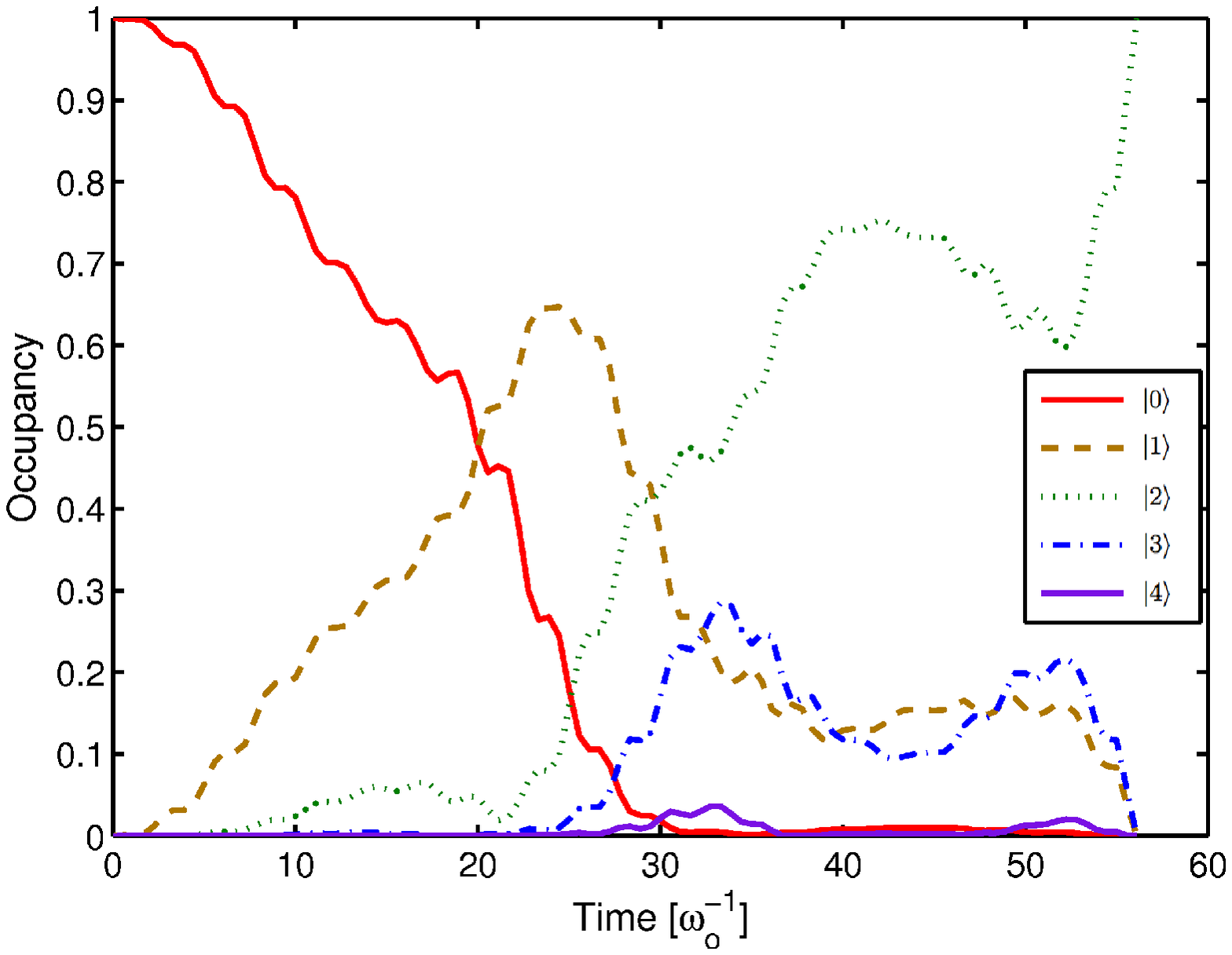}
\caption{\label{fig:pops}Populations of the different states during the pulse taking $|0\rangle$ into $|2\rangle$. It is recognized, that $|0\rangle$ is first excited into $|1\rangle$ and that this one then goes into $|2\rangle$ whereas higher states are somewhat occupied during the pulse and then go to zero. Additional fast modulation is due to counter-rotating terms. }
\end{figure}

The state dynamics during the pulse can be seen in Fig. \ref{fig:pops}.
This analysis confirms the picture of sequential occupation of Fock states, climbing an energy ladder, with some intermediate leakage to higher state that corrects itself in the end. Next to these long-scale Rabi dynamics, there is a fast modulation on the scale of the average driving frequency typical for strong driving where the counter-rotating component of the drive that is not part of standard Rabi physics \cite{Cohen92} becomes important. In the end of the pulse, populations of $|1\rangle$ and $|3\rangle$ are brought back to $|2\rangle$.

\section{Discussion}

\begin{figure}
\includegraphics[width=15cm]{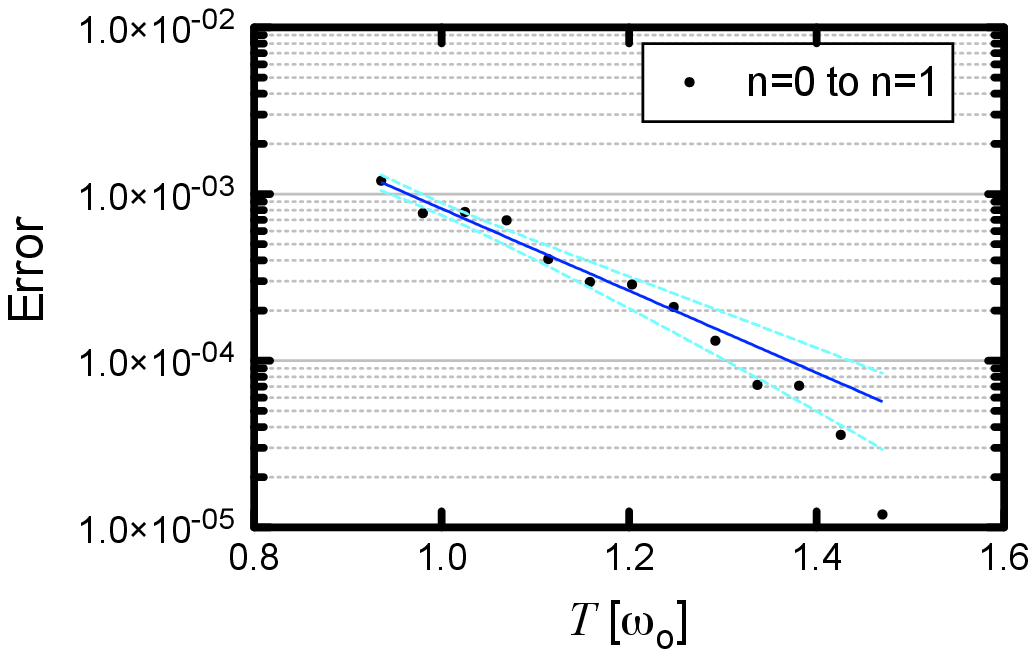}
\caption{\label{fig:fidelity_time} Error for a pulse aiming at creating the first excited state for $\delta=0.12$ along with an exponential fit (solid) and $95\%$ confidence lines (dashed).} 
\end{figure}

\begin{figure}
\includegraphics[width=15cm]{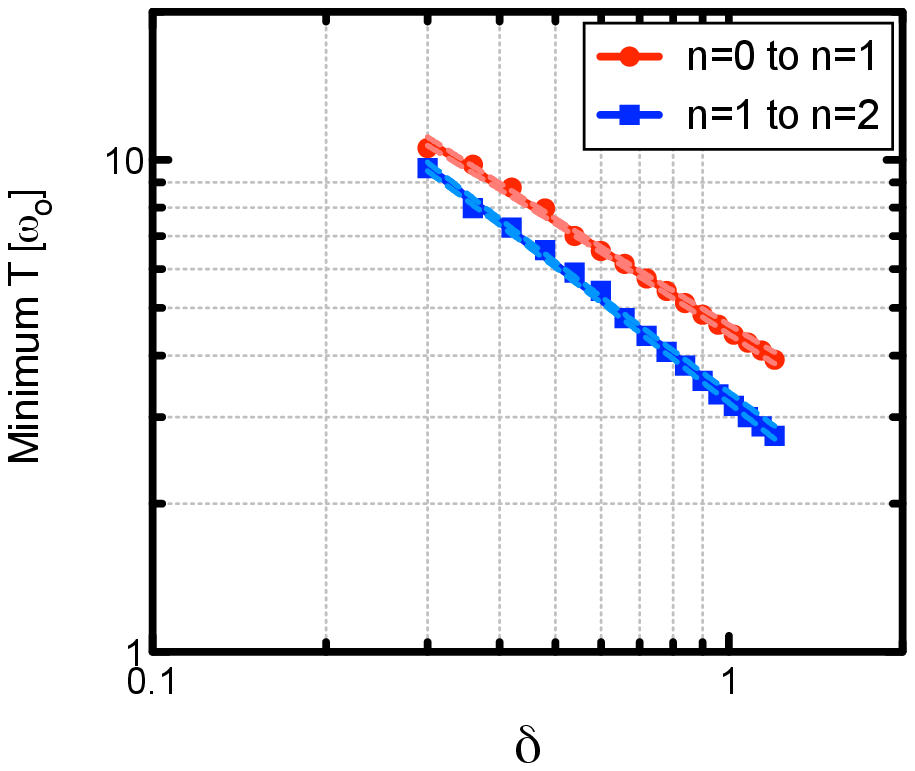}
\caption{\label{fig:scaling} Scaling of the minimal gate time with the nonlinearity parameter with power-law fits. Exponents are $-0.73$ for $|0\rangle\rightarrow |1\rangle$ and $-0.90$ for $|1\rangle\rightarrow|2\rangle$.}
\end{figure}

As mentioned in the introduction, nonlinearity is a resource for the creation of Fock states. At very short pulse durations $t_g$ state transfer pulses must scale with a constant area, i.e., their amplitudes will scale as $1/t_g$. Thus, if the pulse is too short the driving amplitude will be so high that the $\delta$-term in the Hamiltonian can be neglected and Fock state preparation will not be possible. We have numerically investigated the gate fidelity as a function of pulse duration and found the error to be roughly exponentially growing at short times, see fig. \ref{fig:fidelity_time}.
We now investigate the minimal time to reach a fidelity of $0.99999\%$ for pulses $|0\rangle\rightarrow |1\rangle$ and $|1\rangle\rightarrow |2\rangle$, fig. \ref{fig:scaling}. We see that this minimal time is a power law of the nonlinearity parameter $\delta$; $t_{\rm min}\propto \delta^{\alpha}$ with $\alpha_{01}=-0.73 \pm 0.029$ and $\alpha_{12}=-0.90 \pm 0.031$ 

As a reference, we can construct the corresponding power law for non-optimized, single frequency Rabi pulses, starting from the $|0\rangle\rightarrow |1\rangle$ transfer. Transforming to the frame rotating with the driving frequency, the effect of the pulse on the unwanted 
$|1\rangle\rightarrow|2\rangle$ transition is to switch a transition matrix element on and off. This matrix element has to be compared to the detuning between the transition
$|\omega_{21}-\omega_{10}|\simeq\delta$. In order to create a Rabi pulse of fixed
area $\pi$, the maximal amplitude of the transition matrix elements scales as $1/t_g$, and its rate of change thus scales as $1/t_g^2$. In order to not occupy the $|2\rangle$-state in the end of the pulse, the dynamics of the $|1\rangle\leftrightarrow|2\rangle$-transition must be fully adiabatic \cite{Motzoi09}, i.e., following a standard result from Landau-Zener theory
\cite{Zener32,Landau32,Stueckelberg32}, we demand $1/t_g^2\ll \delta^2$ leading to a minimal gate time of $t_g\propto 1/\delta$. We would not expect this scaling law to change for the $0\rightarrow 2$ transition, which is ultimately a sequence of two transitions of length $1/t_g$, only a different prefactor. 

Thus, we can conclude that our optimal pulses {\em qualitatively} extend the limits of Fock state preparation by changing the minimal time to a softer power law, from $1/\delta$ to $1/\delta^{0.73\pm 0.029}$. One can conclude that this is due to quantum interference in the higher levels. The difference is made possible by temporarily occupying higher states and unpopulating them in the course of the pulse. It needs to be remarked, that our pulses qualitatively differ from quantum gate pulses, i.e., rotations of the full basis instead of changing state \cite{Rebentrost09,Motzoi09}, where a $1/\delta$ scaling of the minimal time is found. This suggests that state preparation, which is less constrained, can use quantum interference more efficiently. A related paper has been posted recently \cite{Jirari09} that looks at preparation of $|1\rangle$ exclusively and finds different envelopes, probably due to additional constraints. 

We acknowledge useful discussions with W.A. Coish and L. Hu. Work supported by NSERC discovery grants, quantumworks, and SHARCNET. JMG was supported by CIFAR, MRI, MITACS, NSERC, and DARPA.

\bibliography{frankslibrary,frankspapers}

\end{document}